\documentclass[11pt,a4paper]{article}

\usepackage{jheppub}

\usepackage{mathtools}

\newcommand*{\longhookrightarrow}{\ensuremath{\lhook\joinrel\relbar\joinrel\relbar\joinrel\rightarrow}}

\title{The  \boldmath $\mathfrak{so}$-Kazama-Suzuki Models at Large Level}
\author{Kevin Ferreira}
\author{and Matthias R.\ Gaberdiel}

\affiliation{Institut f\"ur Theoretische Physik, ETH Z\"urich,\\
CH-8093 Z\"urich, Switzerland}

\emailAdd{gaberdiel@itp.phys.ethz.ch}
\emailAdd{kferreira@itp.phys.ethz.ch}

\abstract{The large level limit of the $\mathcal{N}=2$ ${\rm SO}(2N)$ Kazama-Suzuki coset models
is argued to be equivalent to the orbifold of $4N$ free fermions and bosons by the Lie group
${\rm SO}(2N) \times {\rm SO}(2)$. In particular, it is shown that the untwisted sector of the continuous orbifold
accounts for a certain closed subsector of the coset theory. Furthermore, the ground states of the twisted
sectors are identified with specific coset representations, and this identification is checked by
various independent arguments.}

\begin{document}
\maketitle
\flushbottom

\section{Introduction}

In the context of AdS$_3$/CFT$_2$ vector-like dualities and their relation to AdS$_3$ string dualities, 
a link between a Vasiliev higher spin theory on AdS$_3$ \cite{VAS} and the tensionless limit of string 
theory on AdS$_3$ $\times {\rm S}^3\times\mathbb{T}^4$ was recently proposed in \cite{HS}. 
Higher spin/CFT dualities have the advantage of being simpler than their full stringy versions, while
still retaining most of the key features. One may therefore hope that they can provide 
a glimpse of the mechanisms underlying the duality. 
\smallskip

A little while ago \cite{LAR}, a duality relating a family of $2$-dimensional $\mathcal{N}=4$ supersymmetric coset 
CFTs and supersymmetric higher spin theories on AdS$_3$ was proposed, and further tested 
in \cite{Creutzig:2013tja,Candu:2013fta,CH,Beccaria:2014jra}. It generalises 
the bosonic case of \cite{ADS} and the $\mathcal{N}=2$ case of \cite{CREUT}. The
corresponding large $\mathcal{N}=4$ Wolf space cosets possess the same symmetry as string 
theory on  AdS$_3 \times {\rm  S}^3\times {\rm S}^3\times {\rm S}^1$, where the sizes of the two 
S$^3$'s correspond to the level $k$ and the rank $N$ of the $\mathcal{N}=4$ cosets, respectively.

The analysis of \cite{HS} started then by assuming that the tensionless limit of string theory on 
AdS$_3 \times {\rm S}^3\times\mathbb{T}^4$ is dual to the symmetric product orbifold CFT
\begin{equation}\label{DUAL}
\text{Sym}_{N+1}(\mathbb{T}^4)\equiv(\mathbb{T}^4)^{N+1}/S_{N+1}\ .
\end{equation}
In order to relate this stringy duality to the above higher spin -- CFT correspondence, 
it was then natural to consider
the large level limit $k\rightarrow\infty$  of the latter,
since this corresponds to the situation where one of the two S$^3$'s decompactifies 
and the 
dual geometrical background approaches 
${\rm AdS}_3 \times {\rm S}^3\times\mathbb{R}^3 \times {\rm S}^1$ (which, in many respects, is very similar to 
${\rm AdS}_3 \times {\rm S}^3 \times\mathbb{T}^4$). It was shown
in \cite{HS} that the Wolf space cosets can be described, in this limit, as an orbifold  of the theory of 
$4(N+1)$ free bosons and fermions by the continuous group ${\rm U}(N)$, generalising naturally the 
bosonic analysis of \cite{PAU}. (The resulting theory is therefore 
the natural analogue of the ${\rm SU}(N)$ vector model that was proposed by Klebanov \& Polyakov to 
be dual to a higher spin theory on ${\rm AdS}_4$ \cite{KLEB}.)

It was furthermore shown in  \cite{HS} that the $S_{N+1}$ permutation action in \eqref{DUAL} is induced from 
this ${\rm U}(N)$ action via the embedding $S_{N+1}\subset {\rm U}(N)$. In particular, this implies that 
the untwisted sector of the continuous orbifold --- this can be identified with the perturbative part of the
higher spin theory \cite{Creutzig:2013tja,Candu:2013fta} --- is a closed subsector of the untwisted sector 
of the symmetric product orbifold. This relation therefore mirrors very nicely the expectation that 
higher spin theories describe a closed subsector of string theory at the tensionless point.
\smallskip

As a preparation for the ${\cal N}=4$ analysis,  the large level limit of the 
$\mathcal{N}=2$ $\mathfrak{su}$-Kazama-Suzuki cosets \cite{KS} 
\begin{equation}
\frac{\mathfrak{su}(N+1)_{k+N+1}^{(1)}}{\mathfrak{su}(N)_{k+N+1}^{(1)}\oplus \mathfrak{u}(1)^{(1)}_{N(N+1)(N+k+1)}}
\end{equation}
was studied in \cite{SU} (see also 
\cite{Fredenhagen:2012bw, Restuccia:2013tba} for earlier work and \cite{FREDEN} for a subsequent
analysis). These cosets
play a role in the duality with the $\mathcal{N}=2$ supersymmetric higher spin theory on AdS$_3$ \cite{CREUT, CAN}.
It was shown in \cite{SU} that, in the $k\rightarrow\infty$ limit, they have a description as a 
U($N$) orbifold theory of 2$N$ free fermions and bosons transforming as 
\textbf{N}$\oplus\overline{\textbf{N}}$ under U($N$); this is nicely in line 
with what was described  for the $\mathcal{N}=4$ case above.

Minimal model holography for the bosonic $\mathfrak{so}(2N)$ cosets was developed in \cite{AHN, CFT, EVEN}
and was shown to correspond to a higher spin theory on AdS$_3$ with even spin massless gauge fields 
and real massive scalar fields. It is natural to believe that the $\mathcal{N}=2$ $\mathfrak{so}(2N)$ 
Kazama-Suzuki cosets
\begin{equation}\label{SOc}
\frac{\mathfrak{so}(2N+2)_{k+2N}^{(1)}}{\mathfrak{so}(2N)_{k+2N}^{(1)}\oplus \mathfrak{so}(2)_{k+2N}^{(1)}}
\end{equation}
are also dual to a supersymmetric higher spin theory on AdS$_3$. 
One may furthermore expect that the $k\rightarrow\infty$ limit of these cosets can be described as a 
continuous orbifold of free bosons and fermions. 
In this paper we shall argue that, in the case of (\ref{SOc}), the relevant orbifold 
is that of $4N$ free fermions and bosons in the representation 
\begin{equation}
\textbf{2N}_{+1}\ \oplus\ \textbf{2N}_{-1} \qquad \hbox{of ${\rm SO}(2N)\times {\rm SO}(2)$.}
\end{equation}
As in \cite{SU} and \cite{HS}, the untwisted sector of the continuous orbifold 
can be identified with the $k\rightarrow\infty$ limit of the coset model subsector given by
\begin{equation}\label{PERT}
\mathcal{H}_0=\bigoplus_{\Lambda,p}\mathcal{H}_{(0;\Lambda,p)}\otimes\bar{\mathcal{H}}_{(0;\Lambda^*,-p)}\ ,
\end{equation}
where $\Lambda$ are tensorial representations of $\mathfrak{so}(2N)$,  and $p\in\mathbb{Z}$. The remaining 
coset primaries of the form $(\Lambda_+;\Lambda_-,u)$ with $\Lambda_+\neq 0$ and $u\in\frac{1}{2}\mathbb{Z}$, 
and in particular those describing spinor representations, can then be interpreted in terms of the twisted 
sectors of the continuous orbifold. We shall give various pieces of evidence in favour of these claims. In particular,
we establish a precise dictionary between certain coset primaries and the ground states of the twisted sectors,
see section~\ref{TWISTED} below, and test this identification in detail. While much of the analysis is rather similar 
to that of \cite{SU}, there
are interesting subtleties that arise in the $\mathfrak{so}(2N)$ case, and that we have worked out carefully.
\medskip

The paper is organised as follows. In section \ref{KAZ} the coset models are introduced in detail, and our
conventions are described. In section \ref{sec:UNTWISTED} the coset model subsector \eqref{PERT} is identified with 
the untwisted sector of the continuous orbifold, and it is shown that the partition functions of the two descriptions
match. The twisted sectors are then treated in section \ref{TWISTED}: the twisted sector ground states are identified 
with coset primaries, see eqs.~\eqref{Lmstandard}, \eqref{second}, and \eqref{third}, and this identification is then tested
in detail: in section~\ref{sec:confd} it is shown that the conformal dimension of the coset primary agrees with what
one would expect from the twisted sector viewpoint; in section~\ref{sec:ferspec}, the fermionic excitation
spectrum of the twisted sector ground states is determined using coset techniques, and shown to reproduce the 
prediction from the orbifold viewpoint; and in section~\ref{sec:BPS}, the BPS descendants that are expected
to exist from the orbifold viewpoint are constructed explicitly in the coset language. 
Finally, we conclude in section~\ref{CONCLUSION} with an outlook on future work; there
are altogether three appendices in which some of the more technical material is described.

\section{The $\mathcal{N}=2$ Kazama-Suzuki model}\label{KAZ}

Let us begin by introducing the $\mathcal{N}=2$ superconformal field theories of interest for this paper, the 
Kazama-Suzuki \cite{KS} coset models
\begin{equation}
\frac{\mathfrak{so}(2N+2)_{k+2N}^{(1)}}{\mathfrak{so}(2N)_{k+2N}^{(1)}\oplus \mathfrak{so}(2)_{\kappa}^{(1)}}\cong \frac{\mathfrak{so}(2N+2)_k\oplus \mathfrak{so}(4N)_1}{\mathfrak{so}(2N)_{k+2}\oplus \mathfrak{so}(2)_{\kappa}} \ .
\end{equation}
On the left-hand-side we have written both numerator and denominator in terms of $\mathcal{N}=1$ super 
Kac-Moody algebras, whereas the description on the right-hand-side is in terms of the bosonic algebras that 
can be obtained from the ${\cal N}=1$ algebras upon decoupling the fermions. The resulting $4N$ free fermions
are encoded in the $\mathfrak{so}(4N)_1$ algebra. The level of the $\mathfrak{so}(2)$ denominator factor is 
$\kappa=k+2N$ (see appendix \ref{SELRUL}),\footnote{We have defined the `level' of the 
$\mathfrak{so}(2)$ factor in a slightly non-standard fashion: with this normalisation of the generator, the 
eigenvalue spectrum is half-integer (rather than integer).}
 and the central charge of this conformal field theory equals
\begin{align}
\nonumber c&=\frac{1}{2}\cdot4N+\frac{k\cdot\text{dim}(\mathfrak{so}(2N+2))}{k+2N}-\frac{(k+2)\cdot\text{dim}(\mathfrak{so}(2N))}{k+2N}-1 =\frac{6Nk}{k+2N}\ .
\end{align}
In the limit $k\rightarrow\infty$ the central charge approaches $c\cong6N$, which coincides with the central charge of 4$N$ free fermions and bosons.

\noindent The embedding of the coset algebras is induced from the group embeddings
\begin{align}\label{EMBED}
{\rm SO}(2)  \times {\rm SO}(2N)  \ & \longhookrightarrow \ {\rm SO}(2N+2)\\
\nonumber  (z, v)  \ &  \xmapsto{~~~~~~~~}  \ z \oplus v 
\end{align}
where $z \oplus v$ denotes the block-diagonal matrix, while
\begin{align}\label{EMBED2}
{\rm SO}(2)  \times {\rm SO}(2N)  \ & \longhookrightarrow \ {\rm SO}(4N)\\
\nonumber  (z, v) \ & \xmapsto{~~~~~~~~}  \ z \otimes v\ ,
\end{align}
where $z\otimes v$ is the tensor product of matrices.
The coset representations are labelled by triplets $(\Lambda^+;\Lambda^-,u)$, where $\Lambda^+$ is an integrable 
highest weight of $\mathfrak{so}(2N+2)_k$, $\Lambda^-$ is an integrable highest weight of 
$\mathfrak{so}(2N)_{k+2}$, and $u$ is the $\mathfrak{so}(2)_{\kappa}$ weight (that takes values in 
$u\in\frac{1}{2} \mathbb{Z}$). We shall only consider the 
NS sector of the theory; from the above viewpoint this means that we restrict ourselves to the 
vacuum and vector representation of $\mathfrak{so}(4N)_1$.
The selection rules  are (see appendix \ref{SELRUL})
\begin{align}\label{selrules}
\frac{1}{2}\left(\Lambda_N^++\Lambda_{N+1}^+\right) -u \in \mathbb{Z} \ ,\\
l_i^+-l_{i-1}^-\in\mathbb{Z} \ , & & \text{ for }i=2,\ldots,N+1\ ,
\end{align}
where the $l_i^{\pm}$ are the partition coefficients (see appendix \ref{SO}) corresponding to the representations 
$\Lambda^{\pm}$. The first rule constrains the $\mathfrak{so}(2)$ weight to be integer if the representation 
$\Lambda^+$ is tensorial, and half-integer if $\Lambda^+$ is a spinor representation. On the other hand,
 the second rule 
states that $\Lambda^+$ and $\Lambda^-$ have to be both tensorial or both spinorial representations.

Let us denote by $J$ the outer automorphism of the $\mathfrak{so}(2N)$ affine weights (see 
appendix~\ref{FIELDID}) 
\begin{equation}\label{soauto}
J\left[\Lambda^+_0;\Lambda^+_1,\ldots ,\Lambda^+_{N-1},\Lambda^+_{N}\right]
=\left[\Lambda^+_1;\Lambda^+_0,\Lambda^+_2,\ldots ,\Lambda^+_{N},\Lambda^+_{N-1}\right] \ .
\end{equation}
It follows from the branching of the outer automorphisms that the corresponding field identifications of the coset
states take the form 
\begin{equation}\label{fieldiden}
\left(\Lambda^+;\Lambda^-,u\right)\cong \left(J\Lambda^+;J\Lambda^-,u-(k+2N)\right).
\end{equation}
Note that $u$ is changed by an integer, which is compatible with the selection rules above.

\noindent The conformal dimension of the coset representation $(\Lambda^+;\Lambda^-,u)$ is given by
\begin{equation}\label{hform}
h^k_N(\Lambda^+;\Lambda^-,u)=\frac{C^{(N+1)}(\Lambda^+)}{k+2N}-\frac{C^{(N)}(\Lambda^-)}{k+2N}
-\frac{u^2}{2(k+2N)}+n\ ,
\end{equation}
where $C^{(N)}(\Lambda)$ denotes the Casimir of $\mathfrak{so}(2N)$ in the representation $\Lambda$, 
and $n$ is a half-integer specifying at which level the representation $(\Lambda^-,u)$ appears in $\Lambda^+$. 
On the other hand, its U(1) charge equals (see eq.~\eqref{GENERATOR} in appendix \ref{SELRUL})
\begin{equation}\label{CHARGE}
q^k_N(\Lambda^+;\Lambda^-,u)=\frac{2uN}{k+2N}+s\ ,
\end{equation}
where $s$ is an integer encoding the charge contribution from the descendants. For example, the 
coset state $(v;0,\pm1)$ is an (anti-)chiral primary field, with $v$ denoting the vector representation. Indeed 
its conformal dimension and U(1) charge are
\begin{equation}
(v;0,\pm1):~~~~h=\frac{N}{k+2N}\ ,\qquad q=\pm\frac{2N}{k+2N}\ ,
\end{equation}
where $n=0$ and $s=0$ since the vacuum representation appears in the branching of the vector representation for 
both ${\rm SO}(2)$ weights (see appendix \ref{BRALAW}). Similarly, the state $(0;v,\pm1)$, which has
\begin{equation}
(0;v,\pm 1):~~~~h=\frac{k}{2(k+2N)}\ ,\qquad q=\mp\frac{k}{k+2N}\ ,
\end{equation}
with $n=1/2$ and $s=\mp 1$, is (anti-)chiral primary.
Note that  the coset states of the type $(0;s/c,u)$ and $(s/c;0,u)$, where $s$ and $c$ are the two spinor 
representations of $\mathfrak{so}(2N+2)$ or $\mathfrak{so}(2N)$, are not allowed by the selection rules
(\ref{selrules}). Coset representations containing spinor representations for both numerator and denominator exist, 
and they are of the type $(s/c;s/c,u)$ and $(s/c;c/s,u)$. If we require the denominator representation to appear 
at level $0$ of the numerator representation so that the resulting state is a chiral or anti-chiral primary state, the 
allowed possibilities are 
\begin{align}
\nonumber (s;c,-\tfrac{1}{2}) \ , \qquad (c;s,-\tfrac{1}{2}) \ , \qquad (s;s,+\tfrac{1}{2}) \ , \qquad (c;c,+\tfrac{1}{2}) \ ,
\end{align}
whose conformal dimension and ${\rm U}(1)$  charge equal
\begin{equation}
h=\frac{N}{2(k+2N)}\ , \qquad q=\pm\frac{N}{k+2N} \ ,
\end{equation}
where in each case the sign of the charge agrees with the sign of the $\mathfrak{so}(2)$ weight of the coset 
representation.

\section{The continuous orbifold limit: untwisted sector}\label{sec:UNTWISTED}

We are interested in analysing the 
$k\rightarrow\infty$ limit of these coset models. Based on the experience with \cite{SU} one may expect that 
the coset theory becomes in this limit equivalent to an orbifold of a free field theory of $4N$ bosons and fermions 
by the compact group ${\rm SO}(2N) \times {\rm SO}(2)$, where the bosons and fermions transform as
\begin{equation}\label{freefieldrep}
\textbf{2N}_{+1}\oplus\textbf{2N}_{-1} \ .
\end{equation}
Here \textbf{2N} is the vector representation of ${\rm SO}(2N)$, and the subscript refers to the ${\rm SO}(2)$ charge. 
Furthermore, one may expect that the untwisted sector of this orbifold accounts for the 
$\mathcal{W}$ algebra, as well as the representations corresponding to the multiparticle states obtained from 
$(0;v,\pm1)$, i.e., that the untwisted sector of this orbifold can be written as 
\begin{equation}\label{UNTWISTED}
\mathcal{H}_0=\bigoplus_{\Lambda,p}\mathcal{H}_{(0;\Lambda,p)}\otimes\bar{\mathcal{H}}_{(0;\Lambda^*,-p)} \ ,
\end{equation}
where the sum is taken over the tensorial representations $\Lambda$ only, and $p\in\mathbb{Z}$. We note that for 
the tensorial representations $\Lambda^*=\Lambda$.
In the following we want to give concrete evidence for these statements.
\smallskip

As a zeroth order check we note that in the $k\rightarrow\infty$ limit, the central charge approaches
\begin{equation}
c=\frac{6Nk}{k+2N}\cong 6N\ ,
\end{equation}
which matches indeed with the central charge of the theory of $4N$ free bosons and fermions. Furthermore, 
the coset ground states of $(0;v,\pm2)$ may be identified with the free fermions,
since, for $k\rightarrow \infty$, their conformal dimension and ${\rm U}(1)$ charge becomes
\begin{align}
h(0;v,\pm1)&=\frac{k}{2(k+2N)}\cong\frac{1}{2}\\
q(0;v,\pm1)&=\mp\frac{k}{k+2N}\cong\mp1\ .
\end{align}
As in \cite{SU}, the free bosons can then be identified with the  $\frac{1}{2}$-descendants at $h=1$ of these 
ground states. The untwisted sector of the orbifold consists of the multiparticle states generated from these
free fermions and bosons, subject to the condition that they are singlets with respect to the total left-right-symmetric 
${\rm SO}(2N) \times {\rm SO}(2)$ action. In particular, this leads to the condition that the left- and right-moving
representations in eq.~(\ref{UNTWISTED}) are conjugate to one another.

\subsection{The coset partition function}

In order to be more specific about \eqref{UNTWISTED}, we now compute the partition function corresponding 
to the right-hand-side in the 
$k\rightarrow\infty$ limit. The coset character corresponding to $(0;\Lambda,p)$ is the branching function
$b_{(0;\Lambda,p)}^{N,k}(q)$ that appears in the decomposition 
\begin{equation}\label{COS}
{\rm ch}_0^{2N+2,k}(w(z,v),q)\cdot
\chi(\hat{w}(z,v),q)=\sum_{\Lambda,p}
b_{(0;\Lambda,p)}^{N,k}(q)\cdot 
{\rm ch}_{\Lambda,p}^{2N,k+2}(v,z,q)\ ,
\end{equation}
where ${\rm ch}_0^{2N+2,k}$ is the character of the trivial representation of $\mathfrak{so}(2N+2)_k$, 
${\rm ch}_{\Lambda,p}^{2N,k+2}$ is the character of the representation $(\Lambda,p)$ of 
$\mathfrak{so}(2N)_{k+2}\oplus\mathfrak{so}(2)_{\kappa}$, and $\chi$ is the sum of the characters of the 
vacuum and vector representation of $\mathfrak{so}(4N)_1$. Furthermore, $v_i^{\pm 1}$, 
$i=1,\ldots N$ and $z^{\pm 1}$ are the eigenvalues
of the ${\rm SO}(2N)$ and ${\rm SO}(2)$ matrices, respectively, while $w_j^\pm(z,v)$ and 
$\hat{w}_j^\pm(z,v)$ 
are the induced ${\rm SO}(2N+2)$ and ${\rm SO}(4N)$ eigenvalues under the 
embeddings (\ref{EMBED}) and (\ref{EMBED2}), respectively.

\noindent In the $k\rightarrow\infty$ limit, the affine character of ${\Lambda}$ for $\mathfrak{so}(2N)_k$ equals
\begin{align}
\nonumber {\rm ch}_{{\Lambda}}^{2N,k}(v,q)  \cong \ & \frac{q^{h_{\Lambda}}{\rm ch}_{\Lambda}(v)}{ 
\prod_{i=2}^N\prod_{j=1}^{i-1}\prod_{n>0}\left(1-v_jv_i^{-1}q^n\right)\left(1-v_jv_iq^n\right)} \\
&\quad  \cdot\frac{1}{\left(1-v_j^{-1}v_iq^n\right)\left(1-v_j^{-1}v_i^{-1}q^n\right)\left(1-q^n\right)^N}\ ,
\end{align}
where $h_{\Lambda}$ is the conformal weight of $\Lambda$, 
\begin{equation}
h_{\Lambda}=\frac{C^{(N)}(\Lambda)}{k+2N-2}\cong0\ ,
\end{equation}
while ${\rm ch}_{\Lambda}(v)$ is the finite character of the associated (finite) representation $\Lambda$ of 
$\mathfrak{so}(2N)$. 

In order to deduce from this an expression for the branching functions $b_{(0;\Lambda,p)}$ we 
now use the embedding  (\ref{EMBED}) to write the 
 vacuum character of $\mathfrak{so}(2N+2)_k$ as 
\begin{align}\label{COS1}
\nonumber {\rm ch}_0^{2N+2,k}(w(z,v),q)  \cong \ & 
\frac{1}{\prod_{n>0}(1-q^n)^{N+1}\prod_{l=1}^{N}
\left(1-z v_l^{-1}q^n\right)\left(1-zv_lq^n\right)}\\
\nonumber ~& \cdot\frac{1}{\left(1-z^{-1}v_l^{-1}q^n\right)\left(1-z^{-1}v_lq^n\right)}\\
\nonumber ~ & \cdot\frac{1}{\prod_{i=2}^N\prod_{j=1}^{i-1}\left(1-v_iv_j^{-1}q^n\right)\left(1-v_iv_jq^n\right)}\\
~& \cdot\frac{1}{\left(1-v_i^{-1}v_j^{-1}q^n\right)\left(1-v_i^{-1}v_jq^n\right)} \ , 
\end{align}
where we have used that, under the embedding, we have 
$w_1(z,v)=z$, and $w_{j+1}(z,v)=v_j$ for $j=1,\ldots,N$. On the other hand, the 
$\mathfrak{so}(2N)_{k+2}\oplus\mathfrak{so}(2)_{\kappa}$ character evaluated on the same 
eigenvalues $(v,z)$  equals
\begin{align}\label{COS2}
\nonumber {\rm ch}_{\Lambda,p}^{2N,k+2}(v,z,q)\cong \ & \frac{{\rm ch}_{\Lambda}(v)}{
\prod_{n>0}(1-q^n)^N\prod_{i=2}^N\prod_{j=1}^{i-1}\left(1-v_iv_j^{-1}q^n\right)\left(1-v_iv_jq^n\right)}\\
\nonumber ~& \cdot\frac{1}{\left(1-v_i^{-1}v_j^{-1}q^n\right)\left(1-v_i^{-1}v_j^{-1}q^n\right)} 
\cdot\frac{q^{p^2/2\kappa}z^{p}}{\prod_{n>0}(1-q^n)}\ ,
\end{align}
with the last term being the $\mathfrak{so}(2)_{\kappa}$ character of one free boson, evaluated at $u=p$, and $p^2/2\kappa\cong0$ in the $k\rightarrow\infty$ limit. 
Finally, the character $\chi(\hat{w}(z,v),q)$ of the $4N$ free fermions leads to, 
\begin{align}
\nonumber \chi(\hat{w}(v,z),q)=\prod_{n>0}\prod_{i=1}^N&\left(1+z v_i^{-1}q^{n-1/2}\right)
\left(1+z^{-1}v_iq^{n-1/2}\right)\\
 ~& \ \cdot \left(1+z^{-1}v_i^{-1}q^{n-1/2}\right)\left(1+z v_iq^{n-1/2}\right) \ , 
\end{align}
where we have now used the embedding (\ref{EMBED2}). 

Most of the terms now cancel between the two
sides of \eqref{COS}, and the resulting branching function becomes simply (for $k\rightarrow\infty$)
\begin{equation}
b_{(0;\Lambda,p)}^{N,k}(q)\cong a_{(0;\Lambda,p)}^N(q) \ , 
\end{equation}
where $a_{(0;\Lambda,p)}^N(q)$ is the multiplicity of $z^{p} \, {\rm ch}_{\Lambda}(v)$ in
\begin{align}\label{CHAR}
\sum_{\Lambda,p}a_{(0;\Lambda,p)}^N(q)\, z^{p} \, 
{\rm ch}_{\Lambda}(v)\cong \ & 
\prod_{n>0}\prod_{i=1}^N\frac{\left(1+z v_i^{-1}q^{n-1/2}\right)
\left(1+z^{-1}v_iq^{n-1/2}\right)}{\left(1-z v_i^{-1}q^n\right)\left(1-z^{-1}v_iq^n\right)}\\
& \qquad \quad 
\cdot\frac{\left(1+z^{-1}v_i^{-1}q^{n-1/2}\right)\left(1+z v_iq^{n-1/2}\right)}
{\left(1-z^{-1}v_i^{-1}q^n\right)\left(1-z v_iq^n\right)}\ . \nonumber 
\end{align}
Thus the partition function corresponding to \eqref{UNTWISTED} becomes in this limit
\begin{equation}\label{3.13}
\mathcal{Z}_0=(q\bar{q})^{-\frac{N}{4}}\sum_{\Lambda,p}\left|a^N_{(0;\Lambda,p)}(q)\right|^2,\ 
\end{equation}
where the sum runs over all tensorial representations $\Lambda$ as well as all $p\in\mathbb{Z}$. We have also
used the limiting value of the central charge $c\cong 6N$.

\subsection{The orbifold untwisted sector}

It remains to show that (\ref{3.13}) agrees with the untwisted sector of the ${\rm SO}(2N) \times {\rm SO}(2)$
orbifold. By construction, the latter consists of the singlet states made out of the $4N$ free fermions
and bosons, where the free fields transform as in (\ref{freefieldrep}). Inserting the 
group element $(z,v)\in {\rm SO}(2) \times  {\rm SO}(2N)$ into the free partition function leads to 
\begin{align}\label{UNT}
\nonumber (z,v)\cdot\mathcal{Z}_{\rm free}=(q\bar{q})^{-\frac{N}{4}}\prod_{n>0}\prod_{i=1}^N & 
\frac{\left|1+z v_i^{-1}q^{n-1/2}\right|^2\left| 1+z^{-1}
v_iq^{n-1/2}\right|^2}{\left|1-z v_i^{-1}q^n\right|^2\left|1-z^{-1}v_iq^n\right|^2}\\
~& \cdot\frac{\left|1+z^{-1}v_i^{-1}q^{n-1/2}\right|^2\left|1+z v_iq^{n-1/2}\right|^2}
{\left|1-z^{-1}v_i^{-1}q^n\right|^2\left|1-z v_iq^n\right|^2}\ ,
\end{align}
where the $v_i^{\pm1 }$ with $i=1,\ldots,N$ are the eigenvalues of the ${\rm SO}(2N)$ matrix, 
while $z^{\pm 1}$ are the eigenvalues of the ${\rm SO}(2)$ matrix.  

Next we observe that (\ref{UNT}) is just the charge-conjugate square of \eqref{CHAR}, i.e., 
that \eqref{CHAR} describes the decomposition of the chiral (left- or right-moving part) in terms
of representations of ${\rm SO}(2N) \times {\rm SO}(2)$. The singlets of the left-right-symmetric
combination then come from the terms where the two representations --- one from the left, the other
from the right --- are conjugate to one another, and
in each such case, they appear with multiplicity one. Thus it follows that the untwisted sector of the 
continuous orbifold equals
\begin{equation}
\mathcal{Z}_{\rm U}=(q\bar{q})^{-\frac{N}{4}}\sum_{\Lambda,p}\left|a^N_{(0;\Lambda,p)}(q)\right|^2 \ ,
\end{equation}
and therefore agrees precisely with the partition function obtained from the coset theory in the 
$k\rightarrow\infty$ limit. This establishes eq.~(\ref{UNTWISTED}).

\section{Twisted sectors}\label{TWISTED}

Next we want to identify the twisted sector states with specific coset representations. Since we have accounted
already for all states with $\Lambda^+=0$ in terms of the untwisted sector, we should expect that the twisted
sector states will be associated to representations with $\Lambda^+\neq 0$. 

It follows from general orbifold considerations \cite{ORB} 
that the twisted sectors should be labelled by the conjugacy classes
of the orbifold group, see \cite{PAU} for a discussion in a similar context. For a Lie group, the conjugacy classes
can be labelled by elements of the Cartan torus modulo the identification under the action of the Weyl group, i.e., by 
$\mathbb{T}/\mathcal{W}$. The Cartan torus of ${\rm SO}(2N)$ can be taken to be the set of matrices of the type
\begin{equation}
\text{diag} \bigl(A(\theta_1),\ldots, A(\theta_N)\bigr) \ ,
\end{equation}
where $\theta_i\in\left[-\pi,\pi\right]$ and $A(\theta_i)$ is the ${\rm SO}(2)$ matrix given by
\begin{equation}
A(\theta_i)=
\left(\begin{array}{cc}
\cos(\theta_i) & \sin(\theta_i) \\
-\sin(\theta_i) &\cos(\theta_i)
\end{array}\right)\ .
\end{equation}
For the case at hand, the Cartan torus of the Lie group ${\rm SO}(2N) \times {\rm SO}(2)$ is then given by diag($A(\theta_1),\ldots,A(\theta_N))\otimes A(\theta_{N+1})$, again as a tensor product of matrices.
In the following it will be more convenient
to label these group elements by the twists $\beta_i$, $i=1,\ldots ,N+1$, defined by
\begin{equation}
\beta_i=\frac{\theta_i}{2\pi},~~~~~~\beta_i\in\left[-\frac{1}{2},\frac{1}{2}\right].
\end{equation}
Diagonalising the matrices above and taking the tensor product, the elements of the Cartan torus of 
${\rm SO}(2N) \times {\rm SO}(2)$ are then conjugate to 
\begin{align}
\nonumber \text{diag} & \left(e^{2\pi i(\beta_1+\beta_ {N+1})},e^{-2\pi i(\beta_1-\beta_ {N+1})},\ldots,
e^{2\pi i(\beta_N+\beta_ {N+1})},e^{-2\pi i(\beta_N+\beta_ {N+1})},\right.\\
 & \quad \left.e^{2\pi i(\beta_1-\beta_ {N+1})},e^{-2\pi i(\beta_1+\beta_ {N+1})},\ldots,
 e^{2\pi i(\beta_N-\beta_ {N+1})},e^{-2\pi i(\beta_N+\beta_ {N+1})}\right)\ . 
\end{align}
In the corresponding twisted sector, the $4N$ individual fermions and bosons will therefore be twisted by 
$\pm \alpha_i^\pm$, with 
\begin{equation}\label{ALPHAS}
\alpha_i^{\pm} =\beta_{N+1}\pm\beta_i \ , \qquad \text{for $i=1,\ldots,N$.}
\end{equation}

The Weyl group of ${\rm SO}(2N)\times{\rm SO}(2)$ is the semidirect product 
$S_N\ltimes\mathbb{Z}_2^{N-1}$, where $S_N$ acts by permutations on the $\beta_i$, while the 
$\mathbb{Z}_2^{N-1}$ flip the individual signs. (The overall sign and the sign of $\beta_{N+1}$ are immaterial 
since the twists are of the form $\pm \alpha^\pm_i$.) Dividing out the action of the Weyl group can then
be incorporated by taking the sign of the first $N-1$ $\beta_i$'s to be positive, and by arranging their order so
that 
\begin{equation}\label{ORD}
\beta_1\geq\beta_2\geq \cdots \geq\beta_{N-1}\geq|\beta_N|\geq0 \ ,
\end{equation}
while $\beta_{N+1}$ is unconstrained. We should note that 
$\alpha_N^{\pm}$ takes values in $\left[-1,1\right]$, whereas $\alpha_i^+\in\left[-\frac{1}{2},1\right]$
and $\alpha_i^-\in\left[-1,\frac{1}{2}\right]$ for $i=1,\ldots,N-1$.
\smallskip

After these preparations we can now make a proposal for which coset representations correspond to
twisted sector ground states, and what their corresponding twist is. For a given weight $\Lambda_+$ of 
the form 
\begin{equation}
\Lambda_+ = [\Lambda_1,\ldots, \Lambda_{N-1},\Lambda_N,\Lambda_{N+1}]
\end{equation}
we define, for $m=0,\ldots, N-2$, the weights
\begin{equation}\label{Lmstandard}
\Lambda_-^{(m)} = [ \Lambda_1, \ldots, \Lambda_{m-1},\Lambda_m + \Lambda_{m+1},\Lambda_{m+2},\ldots,
\Lambda_{N+1}] \ , \quad
u^{(m)} =   \sum_{j=m+1}^{N-2} \Lambda_j   +  \frac{1}{2} \bigl( \Lambda_N + \Lambda_{N+1} \bigr)\ ,
\end{equation}
or
\begin{align}\label{Lmstandardhat}
\hat\Lambda_-^{(m)} & = [ \Lambda_1, \ldots, \Lambda_{m-1},\Lambda_m + \Lambda_{m+1},\Lambda_{m+2},\ldots,
\Lambda_{N-1}, \Lambda_{N+1},\Lambda_N] \ , \cr
\hat{u}^{(m)} & = -  \sum_{j=m+1}^{N-2} \Lambda_j  - \frac{1}{2} \bigl( \Lambda_N +  \Lambda_{N+1}\bigr) \ .
\end{align}
One easily checks that both 
$(\Lambda_+;\Lambda^{(m)}_-,u^{(m)})$ and $(\Lambda_+;\hat\Lambda^{(m)}_-,\hat{u}^{(m)})$ 
satisfy the selection rules (\ref{selrules}). 
As we shall argue further below, these states define twisted sector ground states, and their twists are given,
for the case of eq.~(\ref{Lmstandard}), by 
\begin{equation}
\beta_j = \frac{1}{k} \Bigl[ \sum_{l=j}^{N-1} \Lambda_l + \tfrac{1}{2} (\Lambda_N + \Lambda_{N+1}) \Bigr] 
= \frac{l_j^-}{k} \ , \qquad (j=1,\ldots,m) \ , 
\end{equation}
\begin{equation}
\beta_j=\frac{1}{k} \Bigl[ \sum_{l=j+1}^{N-1} \Lambda_l + \tfrac{1}{2} (\Lambda_N + \Lambda_{N+1}) \Bigr] 
= \frac{l_j^-}{k} \ , \qquad (j=m+1,\ldots,N-1) \ , 
\end{equation}
as well as 
\begin{equation}
\beta_{N-1} =  \frac{1}{2k} (\Lambda_N + \Lambda_{N+1}) = \frac{l_{N-1}^-}{k} \ , \qquad
\beta_{N} =  \frac{1}{2k} ( \Lambda_{N+1} - \Lambda_{N}) =  \frac{l_{N}^-}{k} \ ,   \qquad 
\beta_{N+1}  = \frac{u^{(m)}}{k} \ ,
\end{equation}
where the $l_j^-$ are the partition coefficients of $\Lambda_-^{(m)}$, see the definition in 
appendix~\ref{SO}. (For the hatted version, eq.~(\ref{Lmstandardhat}), 
the formulae are the same, except that the  twists $\beta_{N}$ and $\beta_{N+1}$ have the opposite sign.) 
We should note that for 
$\Lambda_+$ to be a weight of $\mathfrak{so}(2N+2)_k$, the Dynkin labels (including the affine label)
must satisfy
\begin{equation}\label{affinecond}
\Lambda_0 + \Lambda_1 + 2 \Bigl( \sum_{j=2}^{N-1} \Lambda_j \Bigr) + \Lambda_N + \Lambda_{N+1} = k \ . 
\end{equation}
Applying the automorphism (\ref{soauto}), if necessary, we may always assume that 
$2 \Lambda_1 \leq  \Lambda_0 + \Lambda_1$, and then (\ref{affinecond}) implies that the above $\beta_i$ 
satisfy (for $m=0,\ldots,N-2$)
\begin{equation}\label{betaord}
\frac{1}{2} \geq \beta_1 \geq \beta_2 \geq \cdots \geq \beta_m \geq \beta_{N+1} \geq 
\beta_{m+1} \geq \cdots \geq \beta_{N-1} \geq |\beta_N| \ . 
\end{equation}
(For the hatted solution, $\beta_{N+1}$ is negative, and in (\ref{betaord}) $\beta_{N+1}$
has to be replaced by $-\beta_{N+1}$.) This therefore accounts for the twists with this ordering of the 
twists. 
\smallskip

The remaining two cases (namely when $|\beta_{N+1}|$ is less than $\beta_{N-1}$), are described by
the solutions
\begin{equation}\label{second}
\Lambda_-^{(N-1)} = [\Lambda_1, \ldots, \Lambda_{N-2},\Lambda_{N-1} +\Lambda_N, \Lambda_{N-1} + \Lambda_{N+1}] 
\ , \qquad u^{(N-1)} = \frac{1}{2} \bigl( \Lambda_N + \Lambda_{N+1} \bigr)
\end{equation}
leading to $\beta_{N-1} \geq \beta_{N+1} \geq |\beta_N|$, and 
\begin{equation}\label{third}
\Lambda_-^{(N)} = [\Lambda_1, \ldots, \Lambda_{N-1},\Lambda_{N-1} + \Lambda_N + \Lambda_{N+1}]\ , \qquad
u ^{(N)}= \frac{1}{2} \bigl(\Lambda_N - \Lambda_{N+1} \bigr) \ ,
\end{equation}
leading to $|\beta_N| \geq |\beta_{N+1}| >0$. In either case we also have 
the corresponding hatted version where the last two Dynkin labels of $\Lambda_-$ are interchanged
and the sign of $u$ (and hence $\beta_{N+1}$) is flipped. Again, these representations satisfy 
the selection rules (\ref{selrules}), and their twists 
are still given by 
\begin{equation}
\beta_j = \frac{l_j^-}{k} \ , \quad (j=1,\ldots, N) \ , \qquad \qquad\beta_{N+1} = \frac{u}{k} \ ,
\end{equation}
again with the appropriate modification in the hatted case. 
\medskip

In the following, we want to give evidence for these claims.
As in \cite{SU} we shall show that the 
conformal dimension of the coset primaries agrees with what one expects from a twisted sector ground
state, and that the fermionic excitation spectrum has the expected structure. Finally,
we shall also see that these coset primaries have the appropriate BPS descendants.

\subsection{Conformal dimensions}\label{sec:confd}

The conformal dimension of the ground state in the $\alpha$-twisted sector equals
\begin{equation}\label{HTWIST}
h(\alpha)= \frac{1}{2}\, \sum_{\sigma=\pm}\sum_{i=1}^N|\alpha_i^\sigma|
=\left\{\begin{array}{cc}
\sum_{i=1}^m\beta_i+|\beta_{N+1}|(N-m) &\,\, \,\, \text{if 
$\beta_m \geq |\beta_{N+1}|\geq \beta_{m+1} \geq |\beta_N|$} \\[.2cm]
\sum_{i=1}^{N-1}\beta_i+|\beta_N| & \text{if $|\beta_{N+1}|\leq |\beta_N|$}\ ,\\
\end{array}\right.
\end{equation}
where we have expressed the twists $\alpha_i^\pm$ in terms of the $\beta$ labels as in 
eq.~(\ref{ALPHAS}).
This formula should now agree with the conformal dimension of 
the corresponding coset primary. For the case where $|\beta_{N+1}| > \beta_{N-1}$, 
the relevant coset primary is given in eq.~(\ref{Lmstandard}) (provided that $\beta_{N+1}$ is 
positive; for $\beta_{N+1}$ negative, the hatted solution in eq.~(\ref{Lmstandardhat}) should be 
considered), and the difference of Casimirs becomes
\begin{align}\label{CASFIRST}
\nonumber\Delta C&=C^{(N+1)}(\Lambda_+)-C^{(N)}(\Lambda_-^{(m)})\\
&=\frac{1}{2}k^2\beta_{N+1}^2+k\left(\sum_{i=1}^m\beta_i+(N-m)\beta_{N+1}\right)\ ,
\end{align}
where we have used eq.~(\ref{Cas}). Since $u^{(m)}= k \beta_{N+1}$, it follows from eq.~(\ref{hform}) 
that the conformal dimension of the coset state equals --- we are using here that $n=0$, as follows from
the analysis of appendix~\ref{BRANCH} 
\begin{align}
\nonumber h\left(\Lambda_+;\Lambda^{(m)}_-,u^{(m)}\right)&=\frac{2\Delta C-(u^{(m)})^2}{2(k+2N)}
=\frac{k\left(\sum_{i=1}^m\beta_i+(N-m)\beta_{N+1}\right)}{k+2N}\\
&\cong \sum_{i=1}^m\beta_i+(N-m)\beta_{N+1}\ ,
\end{align}
where we have taken the $k\rightarrow\infty$ limit in the last step. This then reproduces precisely 
eq.~\eqref{HTWIST} for the case $\beta_{N+1}\geq \beta_{N-1}\geq |\beta_N|$. The analysis for the
hatted version, eq.~(\ref{Lmstandardhat}), is essentially identical.

The other two cases are similar. For $|\beta_N|\leq |\beta_{N+1} |\leq \beta_{N-1}$, we need to use
eq.~(\ref{second}) (or its hatted version if $\beta_{N+1}$ is negative), and the difference of 
Casimirs becomes
\begin{equation}
\Delta C=k\sum_{i=1}^{N-1}\beta_i+\frac{1}{2}k^2\beta_{N+1}^2+k\beta_{N+1}\ ,
\end{equation}
which leads to a conformal dimension of
\begin{align}
h\left(\Lambda_+;\Lambda_-^{(N-1)},u^{(N-1)}\right)\cong \sum_{i=1}^{N-1}\beta_i+\beta_{N+1}\ . 
\end{align}
This then reproduces eq.~\eqref{HTWIST} for the case $m=N-1$.
Finally, for $|\beta_{N+1} |\leq |\beta_N|$, we consider the coset primary \eqref{third} (provided that 
$\beta_N$ is positive; for negative $\beta_N$ the hatted version must be taken) and find
\begin{equation}
\Delta C=k\sum_{i=1}^{N-1}\beta_i+\frac{1}{2}k^2\beta_{N+1}^2+k\beta_N\ ,
\end{equation}
and therefore
\begin{equation}
 h\left(\Lambda_+;\Lambda_-^{(N)},u^{(N)}\right)\cong \sum_{i=1}^{N-1}\beta_i+\beta_N\ ,
\end{equation}
which now reproduces the second case of eq.~\eqref{HTWIST}. As before, the analysis of the 
hatted version of eq.~\eqref{third} is similar. 
We should also mention that the sum of the twists
is in all cases equal to 
\begin{equation}
\sum_{\sigma=\pm}\sum_{i=1}^N  \alpha_i^{\sigma} =2N\, \beta_{N+1}\ .
\end{equation}
Because of \eqref{CHARGE}, and using the formula for $u^{(m)}$, this agrees with the 
${\rm U}(1)$ charge of the coset primary in the limit $k\rightarrow\infty$. Again, this is 
in agreement with what one should expect from the orbifold viewpoint.

\subsection{Fermionic excitation spectrum}\label{sec:ferspec}

As a second consistency check we now want to determine the fermionic excitation spectrum
of these coset primaries, using the fact that we can identify the coset primary $(0;v,\pm1)$ 
with the free fermions, see also \cite{SU}. The fusion of $(\Lambda_+;\Lambda_-,u)$ with $(0;v, \epsilon')$ 
yields
\begin{equation}\label{FUSION}
(0;v,\epsilon')\otimes(\Lambda_+;\Lambda_-,u)=(\Lambda_+;\Lambda_-\otimes v,u+\epsilon')\ , 
\end{equation}
where the tensor product $\Lambda_-\otimes v$ decomposes into SO(2$N$) representations as
\begin{equation}\label{TENSORPROD}
\Lambda_-\otimes v=\bigoplus_{\epsilon=\pm}\bigoplus_{r=1}^{N}\Lambda^{(r,\epsilon)}\ ,
\end{equation}
with 
\begin{align}\label{SUBREP}
\Lambda_j^{(r,\epsilon)} & =\left\{\begin{array}{ll} \Lambda_j+\epsilon \qquad & j=r-1\\ 
\Lambda_j-\epsilon & j=r\\ \Lambda_j & \text{otherwise}\end{array}\right. & r\neq N-1\\[.5cm]
\Lambda_j^{(N-1,\epsilon)} &=\left\{\begin{array}{ll} \Lambda_{N-2}+\epsilon \qquad &  j = N-2 \\
\Lambda_{N-1}-\epsilon & j=N-1\\ 
\Lambda_N-\epsilon & j=N\\ \Lambda_j & \text{otherwise\ ,}\end{array}\right. & 
\end{align}
and we have denoted the Dynkin labels of $\Lambda_-$ by $\Lambda_j$. It is here understood that the terms
where one of the resulting Dynkin labels turns out to be negative are not allowed. 

The conformal dimension of the resulting coset primary $(\Lambda_+;\Lambda^{(r,\epsilon)},u+\epsilon')$ can now
be straightforwardly computed. We are only interested in the change of conformal dimension relative to the 
initial ground state, and this equals
\begin{align}
\nonumber\delta h^{(r,\epsilon)}_{\epsilon'}  = & \
h(\Lambda_+;\Lambda^{(r,\epsilon)},u+\epsilon')-h(\Lambda_+;\Lambda_-,u)\\[.4cm]
\nonumber  = & \ \frac{C^{(N+1)}(\Lambda_+)}{k+2N}
-\frac{C^{(N)}(\Lambda^{(r,\epsilon)})}{k+2N}-\frac{(u+\epsilon')^2}{2(k+2N)} + n\\
\nonumber  &\  - \frac{C^{(N+1)}(\Lambda_+)}{k+2N}+\frac{C^{(N)}(\Lambda_-)}{k+2N}+\frac{u^2}{2(k+2N)}\\[.4cm]
  = &\ \frac{2C^{(N)}(\Lambda_-)-2C^{(N)}(\Lambda^{(r,\epsilon)})-1-2u\epsilon'}{2(k+2N)}+n\ .
\end{align}
The difference of the Casimirs can be calculated again with the usual tools. If $r<N-1$ we find 
\begin{equation}
C^{(N)}(\Lambda_-)-C^{(N)}(\Lambda^{(r,\epsilon)})  = 
\epsilon\left(\sum_{j=r}^{N-2}\Lambda_j+\frac{1}{2}(\Lambda_{N-1}+\Lambda_{N})+N-r\right)-\frac{1}{2} \ ,  \\
\end{equation}
while for $r=N-1$ and $r=N$ we get 
\begin{eqnarray}
C^{(N)}(\Lambda_-)-C^{(N)}(\Lambda^{(N-1,\epsilon)}) & = & 
 \epsilon\left(\frac{1}{2}(\Lambda_{N-1}+\Lambda_N)+\frac{1}{2}(N+1)(N+2)\right)-\frac{1}{2} \ , \\
C^{(N)}(\Lambda_-)-C^{(N)}(\Lambda^{(N,\epsilon)}) & = & 
\epsilon\frac{1}{2}(\Lambda_N-\Lambda_{N-1})-\frac{1}{2}\ ,
\end{eqnarray}
respectively. 
The change in conformal dimension then becomes for $r<N-1$
\begin{align}
\nonumber \delta h^{(r,\epsilon)}_{\epsilon'}&= n+\frac{1}{2(k+2N)}
\left(2\epsilon\sum_{j=r}^{N-2}\Lambda_j+\epsilon(\Lambda_N+\Lambda_{N-1})-2u\epsilon'\right)
+\frac{\epsilon(N-r)-1}{(k+2N)}\\
& \cong n+\frac{1}{2(k+2N)}
\left(2\epsilon\sum_{j=r}^{N-2}\Lambda_j+\epsilon(\Lambda_N+\Lambda_{N-1})-2u\epsilon'\right)\ ,
\end{align}
where the last line is obtained by taking the $k\rightarrow\infty$ limit and discarding the second term of 
the first line since its numerator cannot depend on $k$. For $r=N-1$ and $r=N$ we find similarly
\begin{align}
\delta h^{(N-1,\epsilon)}_{\epsilon'} & \cong n+\frac{1}{2(k+2N)}
\left(\epsilon(\Lambda_N+\Lambda_{N-1})-2u\epsilon'\right)\\
\delta h^{(N,\epsilon,)}_{\epsilon'} & \cong n+\frac{1}{2(k+2N)}
\left(\epsilon(\Lambda_N-\Lambda_{N-1})-2u\epsilon'\right)\ .
\end{align}
Next we note that the selection rules of $\mathfrak{so}(4N)_1$ imply that $n=\frac{1}{2}$. 
We can then combine these three equations into a single expression, 
\begin{equation}\label{CONFDIM}
\delta h^{(r,\epsilon)}_{\epsilon'}\cong \frac{1}{2}+\frac{\epsilon l_{r}-p\epsilon'}{k+2N} 
= \frac{1}{2}-\epsilon'\left(\frac{p+l_{r}-2l_{r}\delta_{\epsilon\epsilon'}}{k+2N}\right)\ ,
\end{equation}
where the $l_j$ are the partition coefficients corresponding to the representation $\Lambda_-$,
and $p=u$. In the second step we have also used that the $\epsilon$ and $\epsilon'$ only take the values $\pm1$. 

We can now apply this formula to the coset primaries \eqref{Lmstandard}, and we find that, in the limit 
$k\rightarrow \infty$, 
\begin{equation}
\delta h^{(r,\epsilon)}_{\epsilon'} \ \cong\ 
\frac{1}{2}-\epsilon'\left(\beta_{N+1}+\beta_{r}-2\beta_{r}\delta_{\epsilon\epsilon'}\right) = 
\frac{1}{2}-\epsilon'\alpha^{\sigma}_{r}\ ,
\end{equation}
where $\sigma=-\epsilon\epsilon'$, and we have again used the relation eq.~(\ref{ALPHAS}). Thus
the fermionic excitation spectrum has precisely the claimed structure. The analysis is essentially 
identical for the other two cases, i.e., eqs.~\eqref{second} and \eqref{third}, as well as the hatted 
versions. We should also mention that since finite excitations only change $l_j$ and $p$ by a finite amount
and hence do not modify the expressions for the twists
in the limit $k\rightarrow\infty$,  these excitations live in the same twisted sector 
as the corresponding ground state; this is again in agreement with the orbifold viewpoint.

\subsection{BPS descendants}\label{sec:BPS}

Finally, we analyse the BPS descendants of these twisted sector ground states. The construction again
parallels what was done in \cite{SU}, so we can be somewhat brief. Let us first 
discuss the BPS descendants of the twisted sector ground states \eqref{Lmstandard}.

In order to obtain the chiral primary descendant we apply the fermionic modes $(0;v,+1)$ 
whose mode number is less than $1/2$ to the ground state \eqref{Lmstandard}; this leads to 
\begin{equation}
\Lambda_-^{(m)\, \text{BPS}} = [\Lambda_1,\ldots, \Lambda_{m-1}, \Lambda_{m} + \Lambda_{m+1} +1 , 
\Lambda_{m+2} , \ldots, \Lambda_{N+1}] \ , \qquad
u^{(m)\, \text{BPS}} = u^{(m)} - m  \ ,
\end{equation}
where $m=1,\ldots,N-2$. This defines a chiral primary field, since the difference of the Casimirs is now 
\begin{align}
\nonumber \Delta C&=\ C^{(N+1)}(\Lambda_+)-C^{(N)}(\Lambda_-^{(m)\, \text{BPS}})\\
\nonumber &=\ C^{(N+1)}(\Lambda_+)-C^{(N)}(\Lambda_-^{(m)})-k\sum_{i=1}^m\beta_i-Nm+\frac{1}{2}m^2\\
\nonumber &=\frac{1}{2}(k\beta_{N+1}-m)^2+N(k\beta_{N+1}-m)\\
&=\frac{1}{2}\left[u^{(m)\, \text{BPS}}\right]^2+ Nu^{(m)\, \text{BPS}}\ ,
\end{align}
so that the conformal dimension equals 
\begin{align}
\nonumber h\left(\Lambda_+;\Lambda^{(m)\, \text{BPS}}_-,u^{(m)\, \text{BPS}}\right)
&=\frac{2\Delta C-\left[u^{(m)\, \text{BPS}}\right]^2}{2(k+2N)} =\frac{Nu^{(m)\, \text{BPS}}}{(k+2N)}\\
&=\frac{1}{2}\, q\left(\Lambda_+;\Lambda^{(m)\, \text{BPS}}_-,u^{(m)\, \text{BPS}} \right) \ , 
\end{align}
thus demonstrating that these coset primaries are indeed chiral primary. The analysis for the anti-chiral
primaries is essentially identical, the only difference being that now 
\begin{equation}
u^{(m)\, \overline{\text{BPS}}} = u^{(m)} + 2N - m \ . 
\end{equation}
The other two cases are similar. The BPS descendant of (\ref{second}) equals
\begin{equation}
\Lambda_-^{(N-1)\, \text{BPS}} =
 [\Lambda_1, \ldots, \Lambda_{N-2},\Lambda_{N-1} +\Lambda_N+1, \Lambda_{N-1} + \Lambda_{N+1}+1] 
\ , 
\end{equation} 
where $u^{(N-1)\, \text{BPS}} = u^{(N-1)} - (N-1) $, whereas for the anti-chiral primary we have instead
$u^{(N-1)\, \overline{\text{BPS}}} = u^{(N-1)} +  (N+1 )$. 

\noindent Finally, the chiral-primary descendant of (\ref{third}) is 
\begin{equation}
\Lambda_-^{(N)\, \text{BPS}} = 
[\Lambda_1, \ldots, \Lambda_{N-1},\Lambda_{N-1} + \Lambda_N + \Lambda_{N+1}+2 ]\ , \qquad
u^{(N)\, \text{BPS}}= u^{(N)} - N \ ,
\end{equation}
while for the anti-chiral primary we have instead $u^{(N)\,  \overline{\text{BPS}}}= u^{(N)} + N$. 

It is interesting that these states are chiral and anti-chiral primary even at finite $N$ and $k$, i.e.,
without taking the limit $k\rightarrow \infty$. This is as in the case studied in \cite{SU}.

\section{Conclusion}\label{CONCLUSION}

In this paper we have argued that the large level limit of the ${\cal N}=2$ 
$\mathfrak{so}(2N)$ Kazama-Suzuki models is 
described by an ${\rm SO}(2N) \times {\rm SO}(2)$ orbifold of 4$N$ free fermions and bosons. 
This is similar to what was found in the ${\cal N}=2$ $\mathfrak{su}(N)$ case in \cite{SU} (see also
\cite{FREDEN}), the ${\cal N}=4$ $\mathfrak{su}(N)$ case in \cite{HS}, as well as the original bosonic 
analysis of \cite{PAU}. 
The untwisted sector of the continuous orbifold is accounted for by the coset primaries associated to 
$(0;\Lambda;p)$, which probably again correspond to the `perturbative' scalar degrees of freedom
of the dual higher spin theory.\footnote{This duality has not yet been studied in detail.} 
We also identified the twisted sector states with specific coset primaries, and gave various pieces
of evidence in favour of this identification. 

The analysis was fairly similar to what was done for the ${\cal N}=2$ $\mathfrak{su}(N)$ case in \cite{SU},
but there were some significant differences in the details, i.e., the structure of the Weyl group, 
the form of the twisted sector ground states, etc. The main motivation for performing this analysis in 
detail is that there is also an $\mathfrak{so}(2N)$ version of the ${\cal N}=4$ Wolf space cosets, and
in order to understand its large level limit it may be a good idea to first get acquainted with the
$\mathfrak{so}(2N)$ specific subtleties in the simpler ${\cal N}=2$ setting. The large level limit of the
$\mathfrak{so}(2N)$ ${\cal N}=4$ Wolf space cosets may again have an interpretation as a subsector of
some string theory, and it would obviously be very interesting to understand this in detail. In particular,
since $S_{N+1}\subset {\rm O}(N)\subset {\rm U}(N)$ one may be tempted to believe that the $\mathfrak{so}(2N)$ 
version will also correspond to the symmetric orbifold of the $\mathbb{T}^4$ theory, but that it accounts
for a larger set of higher spin currents. This question will be studied elsewhere.

\acknowledgments

We thank Constantin Candu, Maximilian Kelm and Carl Vollenweider for useful conversations,
and Maximilian Kelm for a careful reading of the manuscript. This work is largely based
on the Master thesis of one of us (KF). We gratefully acknowledge support of the 
Swiss National Science Foundation, in particular, through the NCCR SwissMAP.

\appendix
\section{Coset basics}

\subsection{$\mathfrak{so}(2N)$ conventions}\label{SO}

Representations of ${\rm SO}(2N)$ are labelled by $N$-tuples of positive integer Dynkin labels 
$\left[\Lambda_1,\ldots,\Lambda_N\right]$ which give the highest weight of the representation 
in terms of the fundamental weights. Changing to an orthonormal basis, the labels 
$\left(l_1,\ldots,l_N\right)$ are called partition coefficients and they are related to the Dynkin labels as
\begin{align}
\nonumber l_i &=\sum_{p=i}^{N-2}\Lambda_p+\frac{1}{2}(\Lambda_{N-1}+\Lambda_N)\ , 
\qquad i = 1,\ldots, N-2 \ ,  \\
l_{N-1} &=\frac{1}{2}(\Lambda_{N-1}+\Lambda_N) \ ,\\
\nonumber l_N &=\frac{1}{2}(\Lambda_N-\Lambda_{N-1})\ .
\end{align}
Conversely the Dynkin labels can be written in terms of the partition coefficients $l_i$ as
\begin{align}
\nonumber \Lambda_i &=l_i-l_{i+1} \ , \qquad i = 1,\ldots, N-1 \ , \\
\Lambda_N &=l_N+l_{N-1}\ .
\end{align}

The representations of ${\rm SO}(2N)$ fall into two distinct classes: tensor representations and spinor representations. 
All tensor representations can be obtained by taking suitable tensor powers of the vector representation 
$v=\left[1,0,0,\ldots,0\right]$. They are characterised by the fact that all the $l_i$'s are integer, 
i.e., that $\Lambda_N+\Lambda_{N-1}$ is even.

On the other hand, representations for which $\Lambda_N+\Lambda_{N-1}$ is odd have half-integer $l_i$'s,
and correspond to the spinorial representations. The simplest spinor representations are conventionally called 
the spinor $s=\left[0,0,\ldots,0,1\right]$ and the conjugate spinor $c=\left[0,0,\ldots,1,0\right]$.\\

\noindent The second Casimir of a highest weight representation $\Lambda$ of ${\rm SO}(2N)$ equals
\begin{equation}
C^N(\Lambda)=\frac{1}{2}\sum_{i=1}^N l_i^2+\sum_{i=1}^N l_i(N-i)\ .
\end{equation}
In terms of the Dynkin labels, this expression can be rewritten as 
\begin{align}
\nonumber C^N(\Lambda) =&\frac{1}{2}\sum_{p=1}^{N-2}p\Lambda_p^2+\sum_{p<q}^{N-2}p\Lambda_p\Lambda_q+\frac{2N-1}{2}\sum_{p=1}^{N-2}p\Lambda_p-\frac{1}{2}\sum_{p=1}^{N-2}p^2\Lambda_p \\
&+\frac{1}{8}(\Lambda_N-\Lambda_{N-1})^2+\frac{1}{8}(\Lambda_N+\Lambda_{N-1})^2(N-1) \label{Cas} \\
\nonumber &+\frac{1}{2}(\Lambda_{N-1}+\Lambda_N)\left(\sum_{p=1}^{N-2}p\Lambda_p+\frac{N(N-1)}{2}\right)\ .
\end{align}
For example, the Casimir of the vector representation equals
\begin{equation}
C^{(N)}(v)=\frac{2N-1}{2}\ ,
\end{equation}
while for the spinor/conjugate representation $s/c$ it is equal to
\begin{equation}
C^{(N)}(s/c)=\frac{N(2N-1)}{8}\ .
\end{equation}

\subsection{The coset theory and the selection rules} \label{SELRUL}

Let us denote by $T$ the $\mathfrak{so}(2)$ generator of the bosonic $\mathfrak{so}(2N+2)_k$
algebra of the numerator that commutes with the generators of the $\mathfrak{so}(2N)$ algebra
of the denominator; its OPE is of the form
\begin{equation}
T(z) \,T(w)  \sim \frac{k}{(z-w)^2}\ . 
\end{equation}
Furthermore, let us denote by $V$ the $\mathfrak{so}(2N)$ generator built from bilinears of the free fermions
that also commutes with the $\mathfrak{so}(2N)$ algebra of the denominator; its OPE may be taken to be
of the form 
\begin{equation}
V(z) \,V(w)  \sim \frac{2N}{(z-w)^2}\ . 
\end{equation}
The $\mathfrak{so}(2)$ of the denominator is then $U= T + V$, and its OPE is of the form
\begin{equation}
U(z) \,U(w)  \sim \frac{ (k+2N)}{(z-w)^2}\ . 
\end{equation}
With this normalisation its eigenvalues are all integers; thus the $\mathfrak{so}(2)$ algebra of the 
denominator is at level $\kappa=k+2N$. Furthermore, we note that the free fermion fields of
the numerator carry charge $\pm 1$ with respect to $U$. 

\noindent The surviving $\mathfrak{u}(1)$ generator of the coset algebra is 
\begin{equation}\label{GENERATOR}
\tilde{U}(z)=\frac{2N}{k+2N}\left(T(z)-\frac{k}{2N}V(z)\right)\ .
\end{equation}
By construction it has a regular OPE with the $U$ generator from the denominator; its normalisation has
been fixed so that 
\begin{equation}
\tilde{U}(z)\tilde{U}(w) \sim \frac{c/3}{(z-w)^2}=\frac{2Nk}{k+2N}\frac{1}{(z-w)^2}\ .
\end{equation}
\medskip

The selection rules of the coset can be computed following the methods of \cite{CAN}. 
Suppose $\Lambda^+$ is a highest weight of 
$\mathfrak{so}(2N+2)_k$,  $\Lambda^-$ a highest weight of 
$\mathfrak{so}(2N)_{k+2}$, and $u$ a highest weight of  
$\mathfrak{so}(2)_{k+2N}$. Then the selection rule is
\begin{equation}
\Lambda^+-\Lambda^--\omega_u\in \mathcal{Q}\ ,
\end{equation}
where $\omega_u$ is the weight of the $\mathfrak{so}(2)$ representation labelled by $u$, interpreted as a 
weight of $\mathfrak{so}(2N+2)$ under the embedding, and $\mathcal{Q}$ is the root lattice of 
$\mathfrak{so}(2N+2)$. Given the construction of the last section, the weight $\omega_u$ can be 
explicitly found to be
\begin{equation}
\omega_u=u\, \varepsilon_1\ .
\end{equation}
The condition above then reads:
\begin{align}
\nonumber &\left(l_1^+- u \right)\varepsilon_1+\sum_{i=2}^{N+1}\left(l_i^+-l_i^-\right)\varepsilon_i\in\mathcal{Q} 
\quad
\Longrightarrow &\left\{\begin{array}{ccc} l_1^+- u\in\mathbb{Z}  \qquad & ~\\
l_i^+-l_i^- \in\mathbb{Z} \qquad &\text{for }i=2,\ldots,N+1.\\ \end{array}\right.
\end{align}
Using $l_1^+=\sum_{i=1}^{N-1}\Lambda^+_i+\frac{1}{2}\Lambda^+_N+\frac{1}{2}\Lambda^+_{N+1}$, the upper line yields
\begin{equation}
\frac{1}{2}\left(\Lambda^+_N+\Lambda^+_{N+1}\right)- u \in\mathbb{Z}
\end{equation}
since all Dynkin labels are positive integers.

\subsection{Field identifications}\label{FIELDID}

Field identifications of coset theories were first analysed in \cite{GEP,Moore:1989yh}; below
we shall follow the account given in \cite{FRAN}. For even $N$, the group of outer automorphisms of 
$\mathfrak{so}$(2$N$) is 
$\mathcal{O}=\mathbb{Z}_2\times\mathbb{Z}_2=\left\{1,a\right\}\times\left\{1,a'\right\}$, where 
\begin{align}
\nonumber a\left[\Lambda_0;\Lambda_1,\ldots, \Lambda_{N-1},\Lambda_N\right] &=\left[\Lambda_1;\Lambda_0,\Lambda_2,\ldots,\Lambda_N,\Lambda_{N-1}\right]\\
a'\left[\Lambda_0;\Lambda_1,\ldots,\Lambda_{N-1},\Lambda_N\right] &=
\left[\Lambda_N;\Lambda_{N-1},\ldots,\Lambda_2,\Lambda_1,\Lambda_0\right]\ ,
\end{align}
while for $N$ odd it is $\mathcal{O}=\mathbb{Z}_4=\left\{1,b,b^2,b^3\right\}$ with 
\begin{equation}
b\left[\Lambda_0;\Lambda_1,\ldots,\Lambda_{N-1},\Lambda_N\right]
=\left[\Lambda_{N-1};\Lambda_N,\Lambda_{N-2},\ldots,\Lambda_1,\Lambda_0\right]\ .
\end{equation}
The coset always involves one algebra of each type. Given the structure of the embedding
of $\mathfrak{so}(2N)$ into $\mathfrak{so}(2N+2)$, an automorphism of the numerator only leads to 
an automorphism of the denominator if the relevant automorphisms are $a$ and $b^2$, respectively (or
vice versa).  Thus the field identification group is $\mathbb{Z}_2$, and its action on the weights is given by 
\begin{equation}
J\left[\Lambda_0;\Lambda_1,\ldots,\Lambda_N,\Lambda_{N+1}\right]=
\left[\Lambda_1;\Lambda_0,\Lambda_2,\ldots,\Lambda_{N+1},\Lambda_N\right]\ .
\end{equation}
The action on the $\mathfrak{so}(2)$ weight $u$ is found by noting that $J$ has order 2, together with the fact 
that $u$ is only defined modulo $\kappa = k+2N$. This finally leads to (\ref{fieldiden}).

\section{Branching rules}\label{BRALAW}

In this section the branching rules of $\mathfrak{so}(2N+2)$ into $\mathfrak{so}(2N) \oplus \mathfrak{so}(2)$ 
are described, using the results of \cite{TSU}. In the following we shall label the relevant representations by their 
partition coefficients $l_i$ that were introduced in appendix~\ref{SO}. 

Let then $l=(l_1\geq l_2\geq \cdots \geq \left|l_{N+1}\right|\geq 0)$ be a partition labelling a representation of 
$\mathfrak{so}(2N+2)$, 
$\bar{l}=(\bar{l}_1\geq \bar{l}_2\geq \cdots\geq \left|\bar{l}_N\right|\geq0)$ a representation of 
$\mathfrak{so}(2N)$, and $u$ (a weight) of an $\mathfrak{so}(2)$ representation. In this section 
we will depart from the usual convention employed elsewhere in the paper, and take $l_{N+1}$ 
and $\bar{l}_N$ to be non-negative coefficients, encoding their signs in $\nu,\nu'\in\{\pm\}$. Then 
what was before denoted $l_{N+1}$ and $\bar{l}_N$ is now represented by $\nu'l_{N+1}$ and 
$\nu \bar{l}_N$, respectively.

For the representation $\bar{l}\times u$ to appear in the branching of $l$, the conditions are 
\begin{align}
\nonumber ~& l_i\geq \bar{l}_i\geq l_{i+2}& \text{for }i=1,\ldots,N-1\ , \\ 
~& l_N\geq\bar{l}_{N}\geq 0\ . & ~
\end{align}
Furthermore $u$ must appear as an exponent of $X$ in the finite series expansion of
\begin{equation}
p(X)=X^{\nu\nu'h_{N+1}}\prod_{i=1}^N\left(\frac{X^{h_i+1}-X^{-h_i-1}}{X-X^{-1}}\right)\ ,
\end{equation}
with the coefficient of $X^{u/2}$ giving the corresponding multiplicity. Here the parameters $h_i$ are defined via
\begin{align}\label{HI}
\nonumber h_1&=l_1-\text{max}(l_2,\bar{l}_1)&\\
h_i&=\text{min}(l_i,\bar{l}_{i-1})-\text{max}(l_{i+1},\bar{l}_i) &\text{for }i=2,\ldots,N\\
\nonumber h_{N+1}&=\text{min}(l_{N+1},\bar{l}_N)\ .&
\end{align}
Note that for $n\geq0$, the term in the above generating function is simply
\begin{equation}
\frac{X^{n+1}-X^{-n-1}}{X-X^{-1}}=X^n+X^{n-2}+ \cdots +X^{2-n}+X^{-n}\ .
\end{equation}

\subsection{The ground states of $(\Lambda_+;\Lambda_-^{(m)},u^{(m)})$}\label{BRANCH}

As a specific application of the above results, we now want to show that 
$(\Lambda_-^{(m)},u^{(m)})$ appears in the branching of $\Lambda_+$. Let us first consider
the case when $m=0,\ldots, N-1$, see eq.~(\ref{Lmstandard}), for which $\nu=\nu'$. Then we obtain for the $h_i$ labels
\begin{align}
\nonumber h_i &=0 & \text{for }i=1,\ldots ,m\\
\nonumber h_j &=\Lambda_j & \text{for $j=m+1,\ldots,N-1$}\\
\nonumber h_N &=\text{min}(\Lambda_N,\Lambda_{N+1})\\
h_{N+1}&= \tfrac{1}{2} | \Lambda_{N+1}-\Lambda_N | \ , 
\end{align}
and therefore the generating function is given by
\begin{align}
\nonumber p(X)= X^{\frac{1}{2} | \Lambda_{N+1}-\Lambda_N |}\, 
\prod_{j=m+1}^{N-1} 
\left(X^{\Lambda_{j}}+\cdots +X^{-\Lambda_{j}}\right)
\left(X^{\text{min}(\Lambda_N,\Lambda_{N+1})}+\cdots+X^{-\text{min}(\Lambda_N,\Lambda_{N+1})}\right) \ .
\end{align}
Writing $| \Lambda_{N+1}-\Lambda_N | = \text{max}(\Lambda_N,\Lambda_{N+1}) - \text{min}(\Lambda_N,\Lambda_{N+1})$, and taking the first term from each bracket, the leading exponent is therefore equal to 
\begin{equation}\label{B6}
\sum_{j=m+1}^{N-1} \Lambda_j + \frac{1}{2} (\Lambda_N + \Lambda_{N+1}) = u^{(m)} \ , 
\end{equation}
thus showing that $(\Lambda_-^{(m)},u^{(m)})$ appears in the branching of $\Lambda_+$. Note that the
analysis for the hatted version, eq.~(\ref{Lmstandardhat}), is essentially identical, except that now the 
exponent of the prefactor has the opposite sign, and we take the last term in each bracket. 

In the other two cases the analysis is similar. For the case of (\ref{second}), i.e., $m=N-1$, the only non-zero 
$h_i$ parameters are $h_N=\text{min}(\Lambda_N,\Lambda_{N+1})$ and $h_{N+1} = \frac{1}{2} | \Lambda_{N+1} - \Lambda_N | $, 
and the relevant polynomial is 
\begin{align}
\nonumber p(X) &= X^{\frac{1}{2} | \Lambda_{N+1}-\Lambda_N | }\, \bigl( X^{\text{min}(\Lambda_N,\Lambda_{N+1})} +\cdots + X^{-\text{min}(\Lambda_N,\Lambda_{N+1})} \bigr) \\
& = X^{\frac{1}{2} (\Lambda_{N}+\Lambda_{N+1})} + \cdots + X^{\frac{1}{2} (-3\cdot\text{min}(\Lambda_N, \Lambda_{N+1})+\text{max}(\Lambda_N,\Lambda_{N+1}))} \ . 
\end{align}
This contains $u^{(N-1)}= \frac{1}{2} (\Lambda_{N}+\Lambda_{N+1})$ as an exponent.
Finally, for the case of  (\ref{third}), i.e., $m=N$, the only non-zero $h_i$ parameter is 
$h_{N+1} = \frac{1}{2} | \Lambda_{N+1} - \Lambda_N |$, and thus the only exponent equals
$\nu' \frac{1}{2} | \Lambda_{N+1} - \Lambda_N | = \frac{1}{2} (\Lambda_{N+1} - \Lambda_N) = u^{(N)}$.

\section{Ground state analysis}\label{GROUNDSTATES}

In this appendix we want to show that the coset primaries $(\Lambda_+;\Lambda_-^{(m)},u^{(m)})$
form indeed twisted sector ground states. To this end we analyse, following \cite{SU}, whether the fusion with
the free fermions (that are described by the coset primary $(0;v,\epsilon')$) raises or lowers the conformal
dimension --- if $(\Lambda_+;\Lambda_-^{(m)},u^{(m)})$ is indeed a ground state, then the conformal
dimension must always increase.

The coset state resulting from the fusion of $(\Lambda_+;\Lambda_-^{(m)},u^{(m)})$ with the fermions 
$(0;v,\epsilon')$ was given in \eqref{FUSION} and specified in \eqref{TENSORPROD} and \eqref{SUBREP}. 
We shall first study (see Section~\ref{app:n0}) the question for which combinations of $r,\epsilon$ and
$\epsilon'$ the resulting coset primary appears again at $n=0$; this is then an important ingredient for the 
analysis of see Section~\ref{app:ground} where the conformal dimension of the fermionic descendant
will be determined.

\subsection{Determining the cases with $n=0$}\label{app:n0}

Let us denote, as before,  the partition coefficients of $\Lambda_+$ and $\Lambda_-^{(m)}$ by 
$l_i$ and $\bar{l}_i$, respectively. We consider the $(r,\epsilon)$ fusion channel of \eqref{TENSORPROD} 
and \eqref{SUBREP}, where $r=1,\ldots, N$ and $\epsilon=\pm$. The effect of this fusion is 
to map $\bar{l}_{r} \mapsto \bar{l}_{r}-\epsilon$, while all  other partition coefficients are unchanged;
in addition $u/2=p \mapsto p+\epsilon'$. In a first step we want to study the question for which choices of 
$r,\epsilon$ and $\epsilon'$ the resulting coset primary appears again at $n=0$. 

We shall work at large $k$, and assume that all Dynkin labels (and hence the partition coefficients) 
are large so that the inequalities of eq.~\eqref{HI} do not change under addition or subtraction by $\epsilon$. In the 
same vein, we shall assume that $\nu$ and $\nu'$ (as defined in the second paragraph of 
appendix~\ref{BRALAW}) remain unchanged. Let us first consider the case that $m\leq N-1$, and furthermore that 
$r\leq N-1$. Then, after the fusion with $(r,\epsilon)$, the parameters $h_i$ of eq.~\eqref{HI} take the values
\begin{align}
\nonumber h_i^{(r,\epsilon)} &=\epsilon \, \delta_{i,r}& \text{for $i=1,\ldots ,m-1$}\\
\nonumber h_i^{(r,\epsilon)} &=h_i - \delta_{\epsilon,-}\, \delta_{i,r} & \text{for $i=m,\ldots,N-1$} \\
\nonumber h_{N}^{(r,\epsilon)} &= h_{N}+(\delta_{\epsilon\mu}-1)\delta_{N,r} & \\
\nonumber h_{N+1}^{(r,\epsilon)} &= h_{N+1}-\delta_{\epsilon\mu}\delta_{N,r} \ , & 
\end{align}
where the $h_i$ are the parameters before the fusion, and we define 
$\mu={\rm sign}(\Lambda_{N+1}-\Lambda_{N})$. For $r=1,\ldots,m-1$, the generating function reads
\begin{align}
\nonumber p(X) = X^{\frac{1}{2} | \Lambda_{N+1} - \Lambda_N | } \, \left(\frac{X^{1+\epsilon}-X^{-1-\epsilon}}{X-X^{-1}}\right)\, 
& \prod_{j=m+1}^{N-1} \left(X^{\Lambda_{j}}+\cdots +X^{-\Lambda_{j}}\right) \\
& \cdot\bigl( X^{\text{min}(\Lambda_N,\Lambda_{N+1})} +\cdots + X^{-\text{min}(\Lambda_N,\Lambda_{N+1})} \bigr)\ .
\end{align}
For $\epsilon=-$, $p(X)=0$ and thus the branching is not allowed. 
For $\epsilon=+$, the first term above is simply $X+X^{-1}$ and therefore, taking the first term from each bracket
in the last product, the polynomial contains the powers $p\pm 1$, where we have used (\ref{B6}). Thus, in 
this case, both values $\epsilon'=\pm$ are allowed.

\noindent For  $r=m,\ldots,N-1$, on the other hand, the generating function is
\begin{align}
\nonumber p(X)=X^{\frac{1}{2} | \Lambda_{N+1} - \Lambda_N | } \,  
& \left( X^{\Lambda_{r} - \delta_{\epsilon,-}}+\cdots +X^{-\Lambda_{r} + \delta_{\epsilon,-}}\right) \,
 \prod_{\stackrel{j=m+1}{j\neq r}}^{N-1} \left(X^{\Lambda_{j}}+\cdots +X^{-\Lambda_{j}}\right) \\
& \cdot\bigl( X^{\text{min}(\Lambda_N,\Lambda_{N+1})} +\cdots + X^{-\text{min}(\Lambda_N,\Lambda_{N+1})} \bigr)\ .
\end{align}
Again, taking the first term in each bracket this now leads to $p - \delta_{\epsilon,-}$, and hence this is 
only allowed for $\epsilon'=\epsilon=-1$. 

\noindent The remaining case for $m\leq N-1$ occurs for $r=N$. 
Now the generating function becomes
\begin{align}
\nonumber p(X) = X^{\frac{1}{2} |\Lambda_{N+1} - \Lambda_N | - \delta_{\epsilon,\mu}} \,  
& \left( X^{\text{min}(\Lambda_N,\Lambda_{N+1}) +\delta_{\epsilon,\mu} - 1}+\cdots +X^{-\text{min}(\Lambda_N,\Lambda_{N+1}) -\delta_{\epsilon,\mu} +1}\right) \\
& \cdot\prod_{j=m+1}^{N-1} \left(X^{\Lambda_{j}}+\cdots +X^{-\Lambda_{j}}\right) \ .
\end{align}
Thus it follows that $\epsilon'=-1$, for either sign of $\epsilon$. 
\smallskip

\noindent Finally, we study the case $m=N$. For $r\leq N-1$ the new parameters $h_i$ are then
\begin{align}
& h_i^{(r,\epsilon)} = \epsilon\, \delta_{i,r}  & \text{for $i=1,\ldots, N-1$} \\
& h_{N+1}^{(r,\epsilon)} = \tfrac{1}{2} | \Lambda_{N+1} - \Lambda_N | \ . & ~
\end{align}
In this case, $\epsilon=-$ is not allowed since then $h_{r}=-1$ and $p(X)=0$. On the other hand, 
for $\epsilon=+$ we have
\begin{equation}
p(X)=X^{\frac{1}{2} ( \Lambda_{N+1}-\Lambda_N ) }\, \bigl( X + X^{-1} \bigr) \ , 
\end{equation}
and therefore the allowed exponents are $p\pm 1$, thus allowing both values $\epsilon'=\pm$. The final
case to study is $m=N$ and $r=N$, for which the new parameters $h_i$ equal
\begin{align}
\nonumber h_i^{(N,\epsilon)} &=0 & \text{for $i=1,\ldots,N-1$}\\
h_N^{(N,\epsilon)} &= \epsilon & h_{N+1}^{(N,\epsilon)} = \tfrac{1}{2} | \Lambda_{N+1} - \Lambda_N | \ .
\end{align}
Again, by the same reason as before, only $\epsilon =+$ is possible (since otherwise $p(X)=0$), and then
again both values of $\epsilon'=\pm$ are allowed. In summary, the cases where we have $n=0$ 
for the fusion product are therefore 
\begin{align}
1\leq r\leq m-1: \qquad & \epsilon = +\ , \ \epsilon'=\pm\ ,  \label{C1} \\
m\leq N-1\ , \ \ m\leq r \leq N-1: \qquad & \epsilon=\epsilon' = -  \ ,  \label{C2} \\
m\leq N-1\ , \ \ r = N: \qquad & \epsilon'=- \ , \ \epsilon = \pm  \ ,  \label{C3} \\
m= N\ , \ \ r = N: \qquad & \epsilon= + \ , \ \epsilon' = \pm  \ . \label{C4} 
\end{align}

\subsection{Twisted sector ground state analysis}\label{app:ground}

With these preparations we can now show that the coset states $(\Lambda_+;\Lambda_-^{(m)},u^{(m)})$ are 
indeed ground states. Upon fusion with the fermions in the channel $(r,\epsilon)$, the change in conformal
dimension equals, see eq.~\eqref{CONFDIM},
\begin{equation}\label{compare}
\delta h^{(r,\epsilon)}_{\epsilon'}\cong n + \epsilon \left( \frac{l_r - p \epsilon\epsilon'}{k+ 2N} \right) \cong 
n-\epsilon'\left(\frac{p+l_r-2l_r\delta_{\epsilon\epsilon'}}{k+2N}\right)\ ,
\end{equation}
where $r=1,\ldots,N$. This change can only be negative if $n=0$. Thus in order to show that all of these
changes are positive, it remains to check that, for each of the cases of \linebreak
eqs.~(\ref{C1}) -- (\ref{C4}), 
the second term is actually positive.

For case (\ref{C1}) we use the first formula in (\ref{compare}) and note that $l_r \geq p$, thus implying that
the bracket is always positive. The same argument also applies to case (\ref{C4}). 

On the other hand, for  case (\ref{C3}) we use the second formula in (\ref{compare}), as well as $p\geq l_r$. 
The same argument also applies to (\ref{C2}). This concludes the proof.

\bibliographystyle{alphadin}
\bibliography{document}

\end{document}